\documentclass[final,5p,times]{elsarticle}
\usepackage{amssymb}
\usepackage{amsmath}
\usepackage{lipsum}

\journal{}

\begin{document}

\begin{frontmatter}



\title{Probing Non-Cold Dark Matter with Modified Emergent Dark Energy}


\author[first]{Jun-Chao Wang}
\author[communication]{Yan-Hong Yao}
\affiliation[first]{organization={School of Physics and Mechanical and Electrical Engineering, Longyan University},
city={Longyan},
state={Fujian},
postcode={364012},
country={China}}
\affiliation[communication]{organization={Institute of Fundamental Physics and Quantum Technology, Department of Physics, School of Physical Science and Technology, Ningbo University},
city={Ningbo},
state={Zhejiang},
postcode={315211},
country={China}}

\begin{abstract}
In the standard $\Lambda$CDM cosmology, dark matter is assumed to be a pressureless cold fluid with $w_{\rm dm}=0$. However, the microscopic nature of dark matter remains unknown, and whether its equation-of-state parameter strictly vanishes deserves observational scrutiny. In this work, we introduce a free dark matter equation-of-state parameter $w_{\rm dm}$ within the Modified Emergent Dark Energy (MEDE) framework, constructing the MEDE+$w_{\rm dm}$ model. We systematically derive its background evolution and linear perturbation equations, and constrain the model parameters using Planck 2018 cosmic microwave background (CMB), DESI DR2 baryon acoustic oscillation (BAO), and three independent Type Ia supernova datasets: Pantheon+, Union3, and DES5YR. Using the CMB + BAO + DES5YR combination, we find a preference for a positive dark matter equation of state, $w_{\rm dm}=0.00128\pm0.00044$, together with a 3$\sigma$ level preference for quintessence-like dark energy evolution, $\alpha=-0.66\pm0.22$. When the local $H_0$ prior is included, the constraint on $w_{\rm dm}$ remains essentially unchanged, whereas $\alpha$ shifts toward the $\Lambda$CDM limit, yielding $\alpha=-0.18\pm0.18$. Bayesian model comparison favors $\Lambda$CDM over MEDE+$w_{\rm dm}$, although the preference is reduced to the weak level after including the local $H_0$ prior. Overall, current observations exhibit a $2.6\sigma$--$3\sigma$ preference for a nonzero $w_{\rm dm}$ at the parameter-posterior level, but this indication does not yet constitute a robust detection of non-cold dark matter.
\end{abstract}







\end{frontmatter}




\section{Introduction}
\label{introduction}

The standard $\Lambda$-cold-dark-matter ($\Lambda$CDM) cosmology, in which the late-time acceleration is driven by a cosmological constant $\Lambda$ and structure formation is governed by pressureless cold dark matter (CDM), has achieved remarkable success in fitting a wide range of observations, including the cosmic microwave background (CMB), baryon acoustic oscillations (BAO), and type Ia supernovae (SN)~\cite{Planck2020,Brout2022,DESI2024}. However, as observational precision continues to improve, persistent cracks have emerged within this paradigm. The most prominent of these is the Hubble tension: the Hubble constant inferred from the Planck CMB data within the $\Lambda$CDM framework differs by more than $5\sigma$ from the value determined by the SH0ES collaboration using the local distance ladder~\cite{Riess2022,Breuval2024}. Moreover, the latest DESI DR2 BAO data, when combined with CMB and SN observations, provide strong evidence ($4.2\sigma$) for dynamical dark energy within the $w_0w_a$CDM framework~\cite{Abdul_Karim_2025}. Together with the long-standing fine-tuning and cosmic coincidence problems, these discrepancies have motivated extensive explorations of new physics beyond $\Lambda$CDM, including dynamical dark energy, interacting dark matter--dark energy models, modified gravity, and other possibilities. See Ref.~\cite{Hu2023} for reviews.

Among dynamical dark energy models, the phenomenological emergent dark energy (PEDE) model proposed by Li and Shafieloo \cite{Li2019} has attracted considerable attention because of its economy and its potential to alleviate the Hubble tension. PEDE assumes that dark energy is negligible in the early Universe and only ``emerges'' at late times. The model introduces no additional free parameters beyond the six parameters of spatially flat $\Lambda$CDM. PEDE provides a significantly better fit to the combined CMB, BAO, and SN data than $\Lambda$CDM \cite{Li2019}; subsequent analyses incorporating the full evolution of linear perturbations confirmed that the tension can be reconciled within the six-parameter space~\cite{Pan2020}. Furthermore, within the PEDE+$M_\nu+N_{\rm eff}$ framework, Ref.~\cite{Yang2021} found indications of a nonzero total neutrino mass, suggesting a possible connection between particle physics and cosmological observations in the emergent dark energy scenario.

Meanwhile, the nature of dark matter also deserves closer scrutiny. Although dark matter (DM) is assumed in most cosmological analyses to be a non-relativistic, pressureless cold fluid that interacts only gravitationally and whose equation-of-state parameter satisfies $w_{\rm dm}\equiv p_{\rm dm}/\rho_{\rm dm}=0$, there is no fundamental theoretical reason to exclude the possibility that $w_{\rm dm}\neq0$. The standard CDM paradigm has long faced several potential challenges on small scales, such as the Missing Satellites Problem, the Cusp-Core Problem, and the Too-Big-To-Fail Problem. These issues have motivated the development of various non-standard dark matter scenarios, including warm dark matter (WDM)~\cite{Viel2018,Nadler2021}, fuzzy dark matter (FDM)~\cite{Ferreira2021,Hui2017}, self-interacting dark matter (SIDM)~\cite{Spergel2000,Tulin2018}, and decaying dark matter~\cite{Vattis2019,Abellan2021}.

To describe deviations from the properties of standard CDM in a unified phenomenological manner, Hu proposed the generalized dark matter (GDM) parameterization framework~\cite{Hu1998}, in which different dark matter properties can be characterized by the equation of state, effective sound speed, and viscosity parameter. As the simplest extension of standard CDM, treating the dark matter equation-of-state parameter $w_{\rm dm}$ as a free parameter and testing observationally whether it deviates from zero provides a phenomenological way for probing the nature of dark matter. Subsequently, numerous studies have investigated the possibility of nonzero $w_{\rm dm}$ within different dark energy frameworks, including $\Lambda+w_{\rm dm}$DM, $w+w_{\rm dm}$DM, and $w_0w_a+w_{\rm dm}$DM~\cite{Zavala2005,Xu2013,Kumar2019,Yao2024,Yao2025,kumar2025evidence,li2025exploring,Jiang2026ngo}. Recently, an analysis based on DESI DR2 data found that, when the dark energy equation of state is assumed to be constant, the dark matter equation of state is favored to be positive, with a nonzero preference at the $2.8\sigma$--$3.3\sigma$ level, whereas this significance decreases to $0.8\sigma$--$1.1\sigma$ under the $w_0w_a+w_{\rm dm}$DM parameterization~\cite{li2025exploring}. Yao et al.~\cite{yao2024observational} were the first to relax $w_{\rm dm}$ in the PEDE model, and found that the data favor a negative $w_{\rm dm}$ at the 95\% confidence level. Li et al.~\cite{li2025revisiting} subsequently revisited the PEDE+$w_{\rm dm}$ model using DESI DR2, Planck, and SN data. For the CMB + DESI + DES5YR combination, they obtained $w_{\rm dm}=-0.00093\pm0.00032$, which deviates from zero at approximately $3\sigma$. The significance decreases to $2\sigma$ when Pantheon+ is adopted instead, indicating a certain sensitivity of this result to the SN dataset. These findings suggest that the significance of a nonzero $w_{\rm dm}$ depends on the assumed dark energy model. However, Bayesian evidence analysis strongly disfavors the PEDE+$w_{\rm dm}$ model relative to $\Lambda$CDM~\cite{li2025revisiting}. This indicates that studying nonzero $w_{\rm dm}$ solely within the PEDE dark energy background is still insufficient, making it necessary to investigate more general dark energy frameworks.

The Modified Emergent Dark Energy (MEDE) model~\cite{benaoum2022modified} introduces an additional free parameter $\alpha$, unifying $\Lambda$CDM ($\alpha=0$), PEDE ($\alpha=1$), and a family of more general dark energy behaviors within a single framework: $\alpha>0$ corresponds to phantom behavior, whereas $\alpha<0$ corresponds to quintessence-like evolution. This naturally raises an important question: within the generalized MEDE framework, in which $\alpha$ is allowed to vary freely, how observational constraints on a nonzero $w_{\rm dm}$ will change, and whether the parameter space jointly extended by $\alpha$ and $w_{\rm dm}$ will introduce new parameter degeneracies and further affect the constraints on $H_0$, the matter density, and structure growth.

To address these questions, we introduce a free dark matter equation-of-state parameter $w_{\rm dm}$ within the MEDE framework and construct the MEDE+$w_{\rm dm}$ model. We jointly constrain the model parameters using Planck 2018 CMB, DESI DR2 BAO, and different type Ia supernova datasets, and further investigate the impact of a local $H_0$ prior on the dark energy and dark matter parameters. In particular, we compare three SN samples, Pantheon+, Union3, and DES5YR, to test the robustness of the preference for nonzero $w_{\rm dm}$ against different SN datasets. Finally, we compare the Bayesian evidence for MEDE+$w_{\rm dm}$ and the standard $\Lambda$CDM model, thereby assessing extended dark-sector models from the complementary perspectives of parameter estimation and model selection.

The remainder of this paper is organized as follows. In Sec.~II, we introduce the background evolution and linear perturbation equations of the MEDE+$w_{\rm dm}$ model. Section~III describes the observational datasets, parameter space, and statistical methodology adopted in our analysis. In Sec.~IV, we present the parameter constraints and model comparison results for different data combinations and discuss the effects of different SN datasets and the local $H_0$ prior. Finally, Sec.~V summarizes our main results and presents a discussion.
\section{The MEDE+$w_{\rm dm}$ Model}
\label{model}
In the standard Friedmann--Robertson--Walker (FRW) cosmological model, the Universe is approximated as homogeneous and isotropic on sufficiently large scales. However, observations indicate the presence of small density and spacetime perturbations, which provide the initial conditions for the formation of large-scale structure in the Universe. Therefore, cosmological evolution can generally be described in terms of two components: the homogeneous and isotropic background evolution and linear perturbations.

At the background level, we assume that the Universe consists of radiation, baryons, non-cold dark matter with a constant equation of state, and MEDE dark energy. Under the assumption of spatial flatness, the dimensionless Hubble parameter can be written as
\begin{equation}
\frac{H^2(z)}{H_0^2}
=\Omega_{\rm r}(1+z)^4
+\Omega_{\rm dm}(1+z)^{3(1+w_{\rm dm})}
+\Omega_{\rm b}(1+z)^3
+f_{\rm de}(z),
\label{eq:friedmann}
\end{equation}
where $z$ denotes the redshift and is related to the scale factor $a$ through
$a=1/(1+z)$.
Here, $w_{\rm dm}\equiv p_{\rm dm}/\rho_{\rm dm}$ is the constant dark matter equation-of-state parameter, while $\Omega_{\rm r}$, $\Omega_{\rm dm}$, and $\Omega_{\rm b}$ represent the present-day density parameters of radiation, dark matter, and baryons, respectively. The last term, $f_{\rm de}(z)$, represents the normalized energy density contribution of MEDE dark energy to the Friedmann equation and is given by
\begin{equation}
f_{\rm de}(z)=\Omega_{\rm de}
\left[
1-\tanh\left(
\alpha\log_{10}(1+z)
\right)
\right],
\label{eq:mede_density}
\end{equation}
where $\Omega_{\rm de}$ denotes the present-day dark energy density parameter, and $\alpha$ is an additional parameter characterizing the evolution of dark energy.

Assuming that there are no interactions among the cosmic components, dark energy satisfies an independent energy conservation equation,
\begin{equation}
\dot{\rho}_{\rm de}
+
3H(1+w_{\rm de})\rho_{\rm de}=0,
\label{eq:de_conservation}
\end{equation}
where an overdot denotes differentiation with respect to cosmic time. From the dark energy conservation equation, the dark energy equation of state can be derived as~\cite{benaoum2022modified}
\begin{equation}
w_{\rm de}(z)=-1-
\frac{\alpha}
{3\ln10}
\left[
1+
\tanh\left(
\alpha\log_{10}(1+z)
\right)
\right].
\label{eq:wde}
\end{equation}

It follows from Eq.~\eqref{eq:wde} that $w_{\rm de}$ increases monotonically with cosmic time. For $\alpha>0$, it evolves from $-1-\frac{2\alpha}{3\ln10}$ to $-1$, corresponding to phantom-like dark energy behavior. Conversely, for $\alpha<0$, $w_{\rm de}$ evolves from $-1$ to $-1-\frac{2\alpha}{3\ln10}$, corresponding to quintessence-like dark energy behavior. The cases $\alpha=0$ and $\alpha=1$ reduce to the $\Lambda$CDM and PEDE models, respectively. The equation of state of MEDE dark energy does not cross $-1$.

We further consider linear scalar perturbations in the Universe. We adopt the Newtonian gauge, in which the perturbed FRW metric is written as
\begin{equation}
\mathrm{d}s^2=a^2(\tau)
\left[
-(1+2\psi)d\tau^2
+
(1-2\phi)\delta_{ij}\mathrm{d}x^i\mathrm{d}x^j
\right],
\label{eq:newtonian_metric}
\end{equation}
where $\tau$ is the conformal time, and $\psi$ and $\phi$ denote the two scalar potentials, respectively.

For dark energy and dark matter, the evolution equations for density and velocity perturbations can be obtained from the first-order perturbations of the energy--momentum conservation equations and the Einstein field equations~\cite{kumar2019testing}:
\begin{equation}
\delta'=-(1+w)(\theta-3\phi')
-3\mathcal{H}(c_s^2-w)\delta
-9(1+w)(c_s^2-c_a^2)
\mathcal{H}^2\frac{\theta}{k^2},
\label{eq:delta_perturbation}
\end{equation}
\begin{equation}
\theta'=-(1-3c_s^2)\mathcal{H}\theta
+
\frac{c_s^2}{1+w}k^2\delta
+k^2\psi
-k^2\sigma.
\label{eq:theta_perturbation}
\end{equation}
Here, a prime denotes differentiation with respect to conformal time $\tau$, $\delta\equiv\delta\rho/\rho$ is the relative density perturbation, and $\theta$ is the fluid velocity divergence. The quantity $\mathcal{H}\equiv a'/a=aH$ is the conformal Hubble parameter, $k$ is the comoving wavenumber, $\sigma$ represents the shear perturbation of the fluid, and $w$ denotes the equation-of-state parameter. The adiabatic sound speed is defined as
$
c_a^2
\equiv
p'/\rho'=w-w'/(3\mathcal{H}(1+w))$. The effective sound speed is defined as the ratio of the pressure perturbation to the density perturbation in the fluid rest frame,
$
c_s^2
\equiv
\left.
\frac{\delta p}{\delta\rho}
\right|_{\rm rest}$. Further details can be found in Refs.~\cite{Hu1998}.

For the non-cold dark matter considered in this work, we take  $w_{\rm dm}$ as a constant, and set
\begin{equation}
c_{s,\rm dm}^2=0,
\qquad
\sigma_{\rm dm}=0.
\label{eq:dm_perturbation_parameters}
\end{equation}

For MEDE dark energy, we adopt the time-dependent equation of state $w_{\rm de}$ given by Eq.~\eqref{eq:wde}, together with
\begin{equation}
c_{s,\rm de}^2=1,
\qquad
\sigma_{\rm de}=0.
\label{eq:de_perturbation_parameters}
\end{equation}
The adiabatic sound speed of dark energy is then automatically determined by its time-varying equation of state. With these specifications, the evolution of the MEDE+$w_{\rm dm}$ model is fully determined at both the background and linear perturbation levels.

We now analyze the effects of the dark matter equation-of-state parameter $w_{\rm dm}$ and the dark energy parameter $\alpha$ on the CMB temperature anisotropy power spectrum and the matter power spectrum.

For $w_{\rm dm}>0$, the dark matter density scales as $\rho_{\rm dm}\propto a^{-3(1+w_{\rm dm})}$ and therefore increases more rapidly toward the past, causing matter--radiation equality to occur at an earlier epoch. Consequently, more  modes associated with the acoustic peaks enter the horizon during the matter-dominated era, leading to a suppression of the amplitudes of the acoustic peaks in the CMB TT power spectrum. Meanwhile, the non-adiabatic pressure perturbation causes the gravitational potentials to continue decaying during matter domination. Photons acquire a net energy gain when traversing the decaying gravitational potentials, thereby enhancing the ISW effect and producing an enhancement of power at low $\ell$ in the TT spectrum. At the same time, the decaying potential wells become less effective at confining matter, suppressing the growth of dark matter clustering and consequently reducing $P(k)$. The case of $w_{\rm dm}<0$ exhibits the opposite behavior. The CMB TT and matter power spectra can be found in Fig.~1 of Ref.~\cite{yao2024observational}.
\begin{figure*}
\centering
\includegraphics[width=0.48\textwidth]{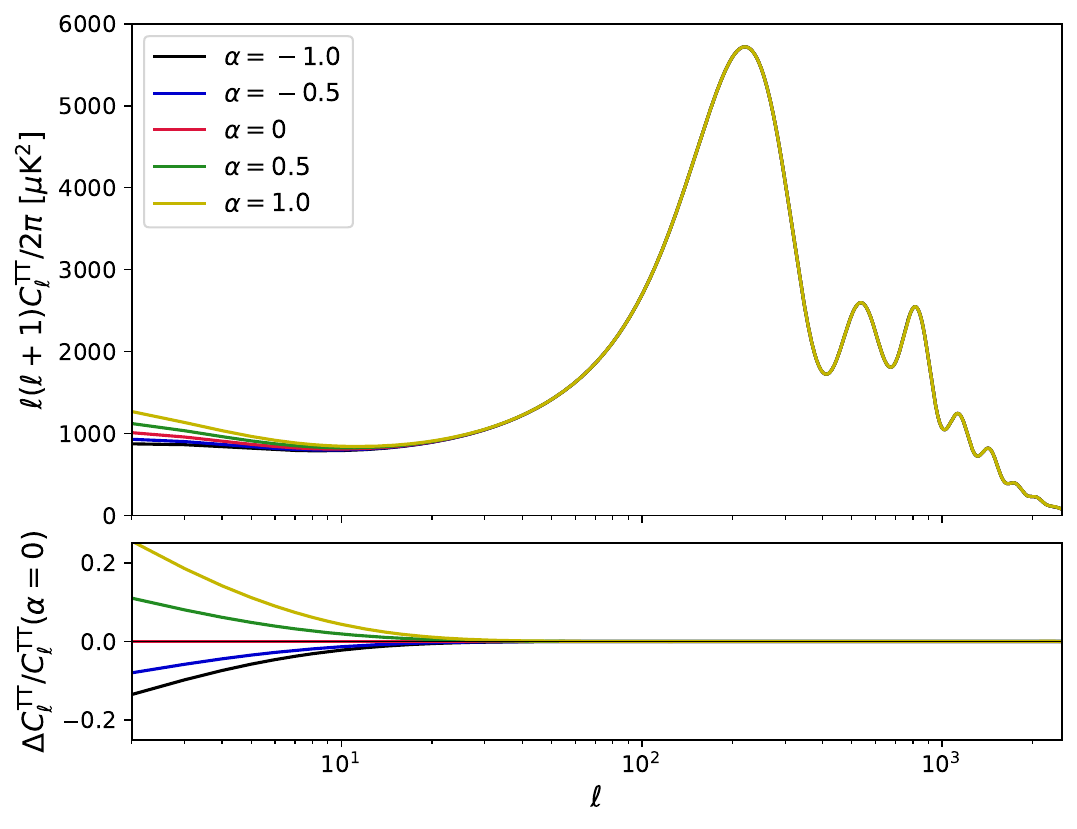}\hfill
\includegraphics[width=0.48\textwidth]{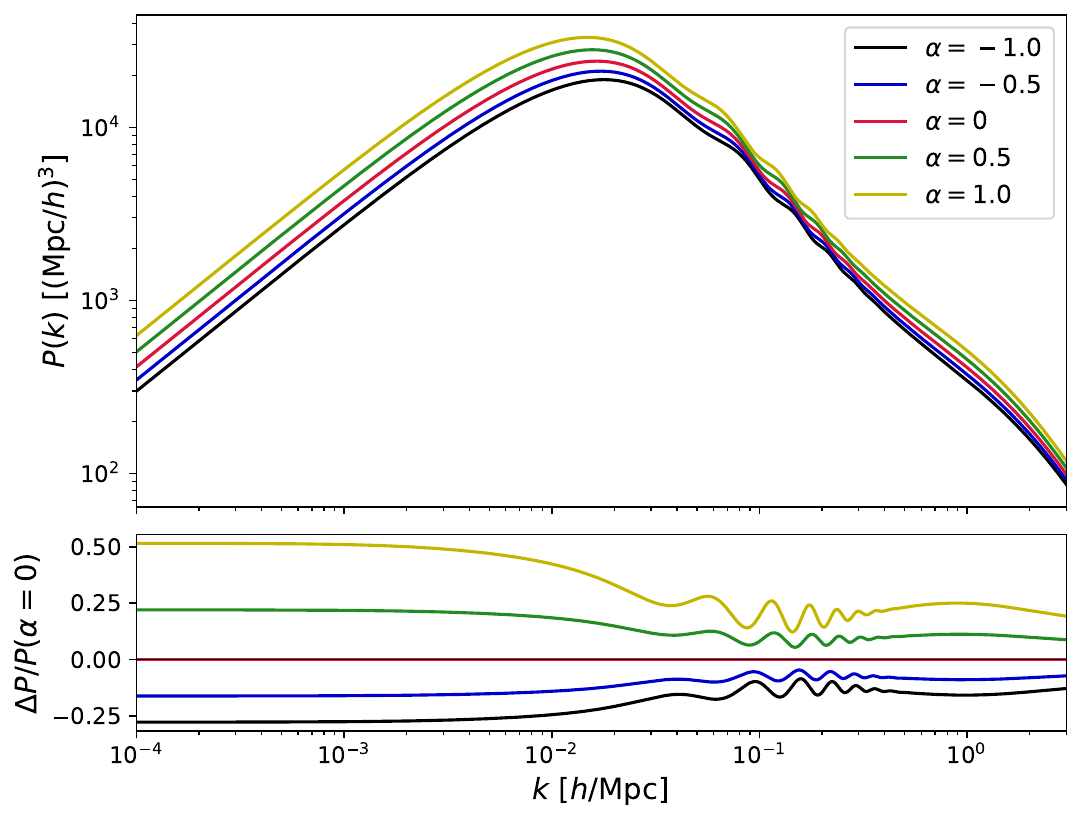}
\caption{CMB temperature power spectra (left panel) and matter power spectra (right panel) for different values of the parameter $\alpha$. The dark matter equation-of-state parameter $w_{\rm dm}$ is fixed to $0$, and the other relevant cosmological parameters are fixed to their mean values extracted from the CMB + BAO + DES5YR data combination listed in Table~\ref{tab:mede_constraints}.}
\label{fig:alpha_spectra}
\end{figure*}

We next turn to the parameter $\alpha$. The MEDE parameter $\alpha$ primarily controls the redshift evolution of the dark energy density, thereby affecting the late-time expansion history of the Universe and the growth of matter density perturbations. Figure~\ref{fig:alpha_spectra} presents the CMB TT power spectra and matter power spectra for different values of $\alpha$.
For $\alpha>0$, dark energy contributes negligibly to the expansion at early times and rapidly becomes dominant only at low redshifts. The matter-dominated epoch is therefore prolonged, allowing density perturbations to undergo more substantial linear growth and increasing the overall amplitude of the matter power spectrum $P(k)$. Meanwhile, the transition from matter domination to dark energy domination is concentrated at low redshifts, causing the gravitational potentials to decay rapidly at recent times. As a result, the late-time ISW effect is enhanced, leading to an upturn of power at low $\ell$ in the TT spectrum. For $\alpha<0$, the opposite occurs: dark energy contributes to the expansion earlier and suppresses matter clustering at an earlier stage, resulting in a reduced $P(k)$. The potential decay is more gradual and occurs earlier, weakening the late-time ISW effect and suppressing the low-$\ell$ TT power.

\section{DATA SETS AND METHODOLOGY}
\label{data}

To constrain the parameters of the MEDE+$w_{\rm dm}$ model and assess the ability of different cosmological models to explain the current observational data, we perform joint constraints on the parameters of the MEDE+$w_{\rm dm}$ and $\Lambda$CDM models using the following observational datasets.

\begin{itemize}
\item \textbf{Cosmic Microwave Background.}

We use the Planck 2018 data release and its corresponding likelihoods. Specifically, we adopt the combined temperature and polarization power-spectrum likelihood, \\$\mathrm{plikTTTEEE+lowl+lowE}$~\cite{Planck2020,aghanim2019planck}, and further include the CMB gravitational lensing likelihood \cite{aghanim2020plancklens}. The CMB data therefore provide important constraints on both the background evolution and the linear perturbation properties of the cosmological model.

\item \textbf{Baryon Acoustic Oscillations.}

For the BAO data, we use the second-year data release of the Dark Energy Spectroscopic Instrument (DESI Year 2)~\cite{Abdul_Karim_2025}. This dataset contains several types of galaxy and quasar samples, including the Bright Galaxy Sample (BGS), Luminous Red Galaxy (LRG) sample, the joint LRG and Emission Line Galaxy (ELG) sample, the ELG sample, the Quasar (QSO) sample, and the Lyman-$\alpha$ (Ly$\alpha$) forest sample.

\item \textbf{Type Ia Supernovae.}

Type Ia supernovae (SNe Ia) are important distance tracers for studying the cosmic expansion history and the nature of dark energy. We consider three independent SNe Ia datasets, Pantheon+~\cite{Brout2022}, Union3~\cite{rubin2025union}, and DES5YR~\cite{abbott2024dark}, in order to examine the impact of different supernova samples on the model constraints.

For the Pantheon+ dataset, the redshift range is $0.01<z<2.26$. We exclude supernovae at $z<0.01$ to reduce the impact of local peculiar velocities on the distance measurements.

For the Union3 dataset, we adopt the 22 data points and covariance matrix provided in Ref.~\cite{kim2024comments}, which are obtained by binning and compressing a sample of 2087 SNe Ia.

For the DES5YR dataset, we use the DES supernova 5YR data, covering the redshift range $0.025<z<1.13$.

\item \textbf{Local $H_0$ Prior.}

We additionally adopt the local measurement of the Hubble constant obtained by Riess et al.~\cite{Breuval2024} using the Hubble Space Telescope (HST) as an $H_0$ prior. Their measurement is $
H_0=(73.17\pm0.86)\ {\rm km\,s^{-1}\,Mpc^{-1}}$.
For the corresponding data combination, we incorporate this measurement as a Gaussian prior in the parameter estimation.

\end{itemize}

For the theoretical calculations, we modify the publicly available Cosmic Linear Anisotropy Solving System (CLASS)\\ code~\cite{blas2011cosmic} by implementing the MEDE+$w_{\rm dm}$ model. This allows us to calculate the background evolution and linear perturbations of the model and to obtain the corresponding theoretical observables.

We perform statistical inference on the model parameters using the Markov Chain Monte Carlo (MCMC) method implemented in MontePython~\cite{550,551}, with random sampling in the parameter space carried out using the Metropolis--Hastings algorithm. To assess the convergence of the MCMC chains, we adopt the Gelman--Rubin convergence criterion~\cite{552} and require $R-1<0.03$. Uniform priors are adopted for all free parameters, with the corresponding parameters and prior ranges listed in Table~\ref{tab:priors}.
\begin{table}[htbp]
\centering
\caption{Uniform priors on the free parameters of the MEDE+$w_{\rm dm}$ and $\Lambda$CDM models.}
\label{tab:priors}
\begin{tabular}{lc}
\hline
Parameter & Prior range \\
\hline
$100\,\Omega_{\rm b}h^2$ & $[0.8,\,2.4]$ \\
$\Omega_{\rm dm}h^2$ & $[0.01,\,0.99]$ \\
$100\,\theta_s$ & $[0.5,\,2]$ \\
$\ln[ 10^{10}A_s]$ & $[1.6,\,4]$ \\
$n_s$ & $[0.8,\,1.2]$ \\
$\tau_{\rm reio}$ & $[0.01,\,0.8]$ \\
$w_{\rm dm}$ & $[-0.1,\,0.1]$ \\
$\alpha$ & $[-5,\,5]$ \\
\hline
\end{tabular}
\end{table}
For a given set of cosmological parameters, the joint likelihood is determined by the likelihoods of the individual observational datasets. To investigate the impact of different SNe Ia datasets on the parameter constraints, we first consider the following three data combinations: CMB + BAO + Pantheon+, CMB + BAO + Union3 and CMB + BAO + DES5YR.
The total likelihood is given by
$-2\ln\mathcal{L}_{\rm tot}=\chi^2_{\rm CMB}
+\chi^2_{\rm BAO}
+\chi^2_{\rm SN}
+\mathrm{const.}
$, where SN denotes the corresponding Pantheon+, Union3, or DES5YR dataset.

We further consider a data combination that includes the local $H_0$ measurement from HST. For this analysis, DES5YR is adopted as the SNe Ia dataset, and we construct the combination CMB + BAO + DES5YR + $H_0$.

We employ the Bayesian evidence to statistically compare the cosmological models. The Bayesian evidence is obtained by integrating the likelihood over the entire parameter space. For a model $M_i$, the Bayesian evidence is defined as
\begin{equation}
B_i
\equiv
P(D|M_i)=
\int
\mathcal{L}(D|\boldsymbol{\theta},M_i)
\pi(\boldsymbol{\theta}|M_i)
{\rm d}\boldsymbol{\theta},
\label{eq:bayesian_evidence}
\end{equation}
where $D$ denotes the observational data, $\mathcal{L}$ is the likelihood function, and $\pi(\boldsymbol{\theta}|M_i)$ is the prior probability distribution of the model parameters. Since the Bayesian evidence integrates over the entire parameter space, unlike the maximum likelihood, it not only accounts for the best-fit quality achievable by a model but also evaluates the parameter space allowed by the model as a whole.

Model comparison is commonly performed using the logarithmic form $\ln B_{ij}=
\ln B_i-\ln B_j$. In this work, we take $i=\Lambda\mathrm{CDM}$ and $j=\mathrm{MEDE}+w_{\rm dm}$. When $\ln B_{ij}>0$, the data favor model $M_i$ over model $M_j$; conversely, $\ln B_{ij}<0$ indicates a preference for model $M_j$.

We use MCEvidence \cite{556,557} to estimate the Bayesian evidence from the converged MCMC chains. This method utilizes the sampling of the posterior distribution obtained from the MCMC chains together with the prior information on the model parameters to estimate the model evidence. 

Finally, we adopt the revised Jeffreys scale \cite{558} to provide a qualitative interpretation of the magnitude of $\ln B_{ij}$. The corresponding criteria are summarized in Table~\ref{tab:jeffreys}.

\begin{table}[htbp]
\centering
\caption{Qualitative interpretation of the Bayesian evidence according to the revised Jeffreys scale.}
\label{tab:jeffreys}
\begin{tabular}{c c}
\hline
$\ln B_{ij}$ & Evidence strength of $M_i$ relative to $M_j$ \
\\
\hline
$0\leq\ln B_{ij}<1$ & Weak \
\\
$1\leq\ln B_{ij}<3$ & Positive \
\\
$3\leq\ln B_{ij}<5$ & Strong \
\\
$\ln B_{ij}\geq5$ & Very strong \
\\
\hline
\end{tabular}
\end{table}

Therefore, by comparing $\ln B_{\Lambda{\rm CDM},,{\rm MEDE}+w_{\rm dm}}$, we can quantitatively assess the relative support provided by the observational data for the $\Lambda$CDM and MEDE+$w_{\rm dm}$ models under a common choice of datasets and prior assumptions.
\section{RESULTS}
\label{results}

Table~\ref{tab:mede_constraints} presents the cosmological parameter constraints on the MEDE+$w_{\rm dm}$ model obtained from the CMB + BAO + Pantheon+, CMB + BAO + Union3, CMB + BAO + DES5YR, and CMB + BAO + DES5YR + $H_0$ data combinations. We first discuss the impact of different supernova samples on the parameter constraints of the MEDE+$w_{\rm dm}$ model and then examine the changes induced by including the local $H_0$ prior.

We begin with the first three data combinations, which do not include the $H_0$ prior. The standard cosmological parameters constrained by combining CMB + BAO with the three supernova samples, Pantheon+, Union3, and DES5YR, differ by less than $1\sigma$ and are therefore highly consistent with one another. For all three SN datasets, we obtain a positive $w_{\rm dm}$, corresponding to a deviation of approximately $2.6\sigma$--$3\sigma$ from $w_{\rm dm}=0$ in the standard cold dark matter model. A positive $w_{\rm dm}$ implies that the dark matter density evolves slightly faster than in standard CDM. This additional degree of freedom further introduces degeneracies with the matter density and other cosmological parameters, thereby affecting the joint parameter constraints through the CMB and BAO constraints on the background evolution and distance scales. For the dark energy parameter $\alpha$, all three supernova samples favor negative values, with deviations from $\alpha=0$ at the $2\sigma$--$3\sigma$ level. Among them, DES5YR yields the most pronounced preference for quintessence-like dark energy behavior.
\begin{figure}[htbp]
\centering
\includegraphics[width=0.4\textwidth]{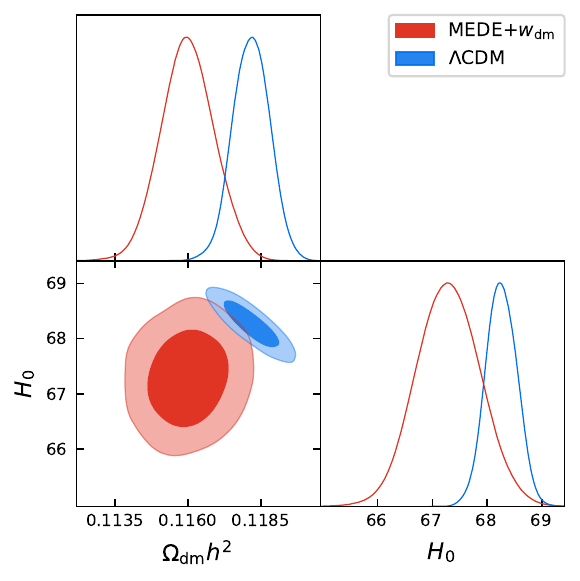}
\caption{Comparison of the posterior constraints on $\Omega_{\rm dm}h^2$ and $H_0$ between the $\Lambda$CDM and MEDE+$w_{\rm dm}$ models using the CMB + BAO + DES5YR data combination. The one-dimensional marginalized posterior distributions and two-dimensional joint contours at 68\% and 95\% CL are shown.}
\label{fig:triangle_omega_dm_H0}
\end{figure}

Figure~\ref{fig:triangle_omega_dm_H0} compares the constraints on $\Omega_{\rm dm}h^2$ and $H_0$ from the $\Lambda$CDM and MEDE+$w_{\rm dm}$ models for the CMB + BAO + DES5YR data combination, where $h\equiv H_0/(100,{\rm km,s^{-1},Mpc^{-1}})$ is the dimensionless Hubble parameter. Compared with $\Lambda$CDM, the MEDE+$w_{\rm dm}$ model favors lower values of $\Omega_{\rm dm}h^2$ and $H_0$. The lower value of $H_0$ is related to the modification of the late-time expansion history induced by negative $\alpha$ and its parameter degeneracy with $H_0$. On the other hand, a positive $w_{\rm dm}$ modifies the evolution of the dark matter energy density in the early Universe, increasing its early-time energy density for a fixed present-day density. This effect can be compensated by shifting $\Omega_{\rm dm}h^2$ to lower values, thereby maintaining a good fit to the early-Universe observables and the acoustic scale.
\begin{figure}[htbp]
\centering
\includegraphics[width=\columnwidth]{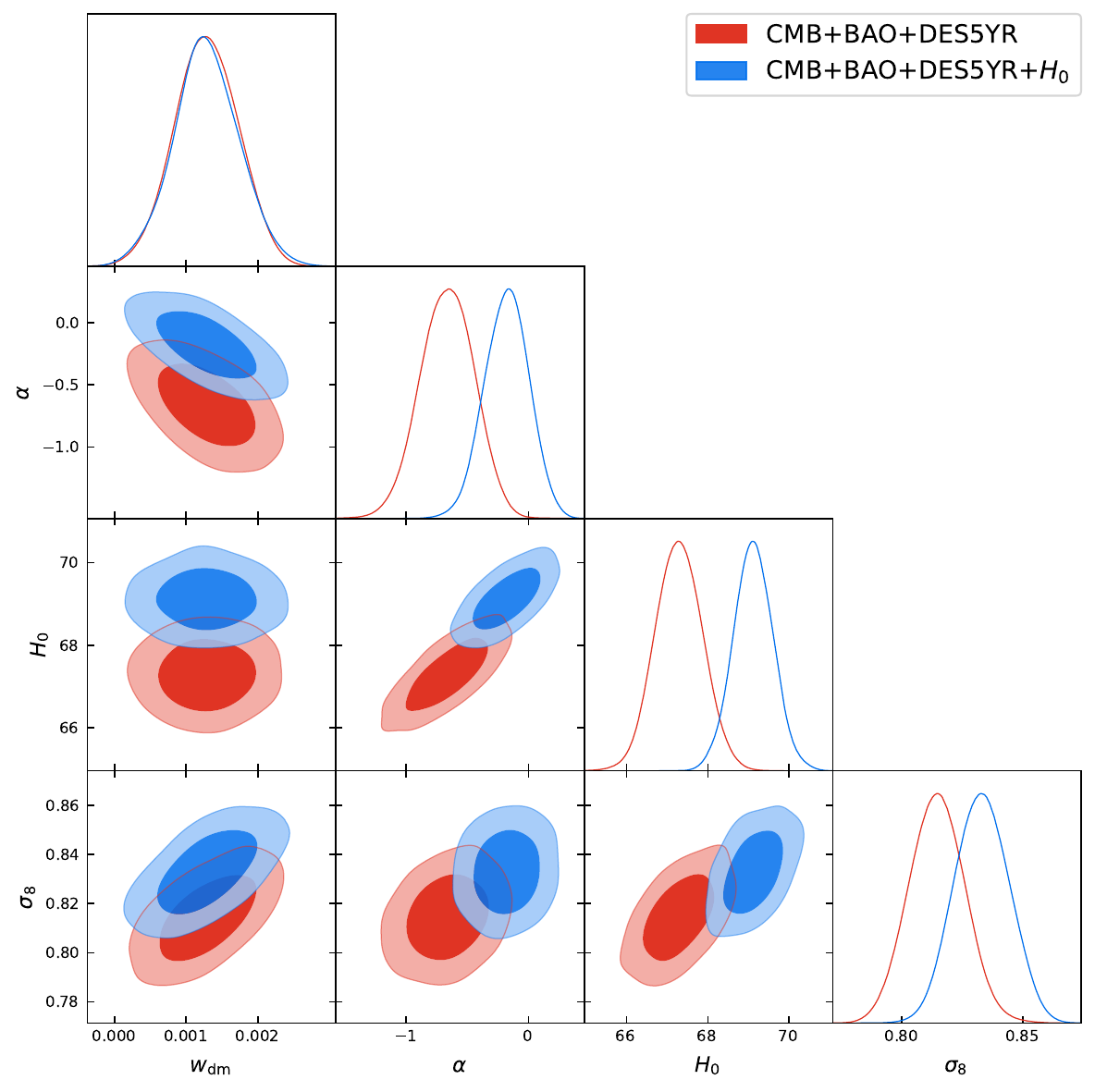}
\caption{Two-dimensional parameter constraints on $w_{\rm dm}$, $\alpha$, $H_0$ and $\sigma_8$ in the MEDE+$w_{\rm dm}$ model, obtained from the CMB + BAO + DES5YR and CMB + BAO + DES5YR + $H_0$ data combinations. The one-dimensional marginalized posterior distributions and two-dimensional joint contours at 68\% and 95\% CL are shown.}
\label{fig:mede_alpha_wdm_h0_sigma8}
\end{figure}
We next investigate the impact of including the local $H_0$ prior. Figure~\ref{fig:mede_alpha_wdm_h0_sigma8} displays the two-dimensional parameter constraints on $w_{\rm dm}$, $\alpha$,  $H_0$, and $\sigma_8$ for the CMB + BAO + DES5YR and CMB + BAO + DES5YR + $H_0$ data combinations. The high-$H_0$ prior has a noticeable impact on the dark energy parameter $\alpha$ in the MEDE+$w_{\rm dm}$ model. The mean value of $\alpha$ shifts significantly toward zero, substantially weakening the preference for quintessence-like dark energy behavior previously obtained from DES5YR. At the $1\sigma$ level, $\alpha=0$ becomes consistent with the data. This suggests that the high-$H_0$ prior drives the MEDE+$w_{\rm dm}$ model toward dark energy behavior closer to the $\Lambda$CDM limit.

Another noteworthy change is the increase in $\sigma_8$. As illustrated in Figure~\ref{fig:mede_alpha_wdm_h0_sigma8}, $H_0$ and $\sigma_8$ exhibit a clear positive correlation. Consequently, the high-$H_0$ prior shifts the allowed parameter space toward larger values of $H_0$ while simultaneously favoring larger values of $\sigma_8$. This behavior reflects the joint degeneracies among the background expansion history, matter density, and structure growth. By comparison, the other standard cosmological parameters and $w_{\rm dm}$ show no significant changes after including the $H_0$ prior, suggesting that the primary impact of the local $H_0$ measurement is concentrated on the late-time expansion parameters and their correlation with the amplitude of structure growth.

Finally, we compare the MEDE+$w_{\rm dm}$ and $\Lambda$CDM models using Bayesian evidence, with the results listed in the last column of Table~\ref{tab:mede_constraints}. All data combinations yield positive Bayes factors, indicating a preference for $\Lambda$CDM over MEDE+$w_{\rm dm}$ from the perspective of model selection. The Pantheon+ and Union3 data combinations provide strong support for $\Lambda$CDM, whereas the preference for $\Lambda$CDM is weaker for DES5YR. After including the $H_0$ prior, the dark energy parameter of the MEDE+$w_{\rm dm}$ model moves further toward the $\Lambda$CDM limit, and only weak evidence in favor of $\Lambda$CDM remains.
\begin{table*}
\footnotesize
\centering
\caption{The mean values and 68\% CL intervals of the free and derived parameters of the MEDE+$w_{\rm dm}$ model for the CMB + BAO + Pantheon+, CMB + BAO + Union3, CMB + BAO + DES5YR, and CMB + BAO + DES5YR + $H_0$ data combinations. The last row gives the Bayes factor $\ln B_{\Lambda\mathrm{CDM},\mathrm{MEDE}+w_{\rm dm}}$.}
\label{tab:mede_constraints}
\begin{tabular}{lcccc}
\hline
parameters & CMB + BAO + Pantheon+ & CMB + BAO + Union3 & CMB + BAO + DES5YR & CMB + BAO + DES5YR + $H_0$ \\
\hline
$100\,\Omega_{\rm b}h^2$ & $2.238 \pm 0.015$ & $2.237 \pm 0.015$ & $2.237 \pm 0.015$ & $2.239 \pm 0.016$ \\
$\Omega_{\rm dm}h^2$ & $0.11646^{+0.00091}_{-0.00082}$ & $0.11611 \pm 0.00095$ & $0.11599 \pm 0.00091$ & $0.11643 \pm 0.00093$ \\
$100\,\theta_s$ & $1.04198 \pm 0.00028$ & $1.04198 \pm 0.00029$ & $1.04200 \pm 0.00028$ & $1.04197 \pm 0.00028$ \\
$\ln[ 10^{10}A_s]$ & $3.052^{+0.012}_{-0.014}$ & $3.053 \pm 0.015$ & $3.053^{+0.013}_{-0.015}$ & $3.050 \pm 0.015$ \\
$n_s$ & $0.9689 \pm 0.0036$ & $0.9691 \pm 0.0036$ & $0.9693 \pm 0.0038$ & $0.9682 \pm 0.0038$ \\
$\tau_{\rm reio}$ & $0.0570^{+0.0065}_{-0.0073}$ & $0.0571^{+0.0068}_{-0.0079}$ & $0.0576^{+0.0069}_{-0.0079}$ & $0.0556^{+0.0069}_{-0.0078}$ \\
$w_{\rm dm}$ & $0.00110 \pm 0.00042$ & $0.00125 \pm 0.00046$ & $0.00128 \pm 0.00044$ & $0.00128 \pm 0.00046$ \\
$\alpha$ & $-0.42 \pm 0.20$ & $-0.61^{+0.32}_{-0.27}$ & $-0.66 \pm 0.22$ & $-0.18 \pm 0.18$ \\
$H_0$ & $67.90 \pm 0.56$ & $67.46 \pm 0.78$ & $67.29 \pm 0.58$ & $69.12 \pm 0.49$ \\
$\sigma_8$ & $0.819 \pm 0.011$ & $0.816 \pm 0.012$ & $0.815 \pm 0.011$ & $0.833 \pm 0.011$ \\
\hline
$\ln B_{\Lambda\mathrm{CDM},\mathrm{MEDE}+w_{\rm dm}}$ & $4.96$ & $3.77$ & $1.89$ & $0.32$ \\
\hline
\end{tabular}
\end{table*}

\section{CONCLUSION AND DISCUSSION}
\label{conclusion}

The standard $\Lambda$CDM model assumes dark matter to be strictly pressureless cold dark matter, namely $w_{\rm dm}=0$. However, the microscopic nature of dark matter remains unknown, and there is no fundamental theoretical or observational reason requiring its equation of state to be exactly zero. Therefore, treating $w_{\rm dm}$ as a free parameter and testing whether it deviates from zero using cosmological observations provides a simple and effective approach to exploring the nature of dark matter. On the other hand, cosmological constraints on the dark matter equation of state may depend on the assumed dark energy model. As an emergent dark energy model that introduces no additional free parameters, PEDE has been used to investigate the possibility of a nonzero $w_{\rm dm}$. However, its Bayesian evidence analysis showed that cosmological observations strongly favor $\Lambda$CDM, with $\ln B$ reaching 17.93 relative to the PEDE+$w_{\rm dm}$ model~\cite{li2025revisiting}.

PEDE is a special case of the more general MEDE model. By introducing the parameter $\alpha$, MEDE unifies $\Lambda$CDM, PEDE, and more general phantom-like and quintessence-like dark energy evolutions within a single framework. This naturally raises the question of how the constraints on the dark matter equation of state change when the dark energy evolution is no longer fixed to that of PEDE but instead allows $\alpha$ to vary freely. Motivated by this consideration, we construct the MEDE+$w_{\rm dm}$ model and systematically investigate its background and perturbation evolution, as well as the constraints on its parameters from the latest cosmological observations.

Within the MEDE+$w_{\rm dm}$ framework, the current CMB, DESI DR2 BAO, and DES5YR data yield, a preference for a positive dark matter equation of state, $w_{\rm dm}=0.00128\pm 0.00044$, together with quintessence-like dark energy evolution characterized by $\alpha=-0.66\pm0.22$. Since the MEDE dark energy equation of state does not naturally cross $-1$, quintessence-like dark energy remains above $-1$ throughout the entire cosmic history, thereby avoiding the emergence of negative kinetic energy and violation of the null energy condition (NEC). The Bayesian criterion, $\ln B_{ij}=1.89$, indicates positive evidence in favor of $\Lambda$CDM in this case. After including the local $H_0$ prior, the dark energy parameter shifts to $\alpha=-0.18\pm0.18$. Consequently, the original preference for quintessence-like dark energy is substantially weakened, and $\alpha=0$ becomes consistent with the data at the $1\sigma$ level. Meanwhile, the Bayesian criterion decreases to $\ln B_{ij}=0.32$, leaving only weak evidence in favor of $\Lambda$CDM. This suggests that the local $H_0$ measurement tends to drive the dark energy behavior of the MEDE+$w_{\rm dm}$ model closer to the $\Lambda$CDM limit. In contrast, $w_{\rm dm}$ does not change significantly after including the $H_0$ prior, indicating that the direct impact of the local $H_0$ measurement on the dark matter equation of state is relatively limited, with its primary effect concentrated on the late-time expansion history and the associated parameter space.

It is worth noting that the positive $w_{\rm dm}$ obtained in this work is in good agreement with the result of Ref.~\cite{li2025exploring} within the $\Lambda$ WDM framework, suggesting that the preference for a positive constant dark matter equation of state is not driven solely by a particular dark energy model. Although a negative $w_{\rm dm}$ was obtained in the PEDE+$w_{\rm dm}$ model, this model itself is strongly disfavored by the Bayesian evidence~\cite{li2025revisiting} and therefore cannot be regarded as evidence of equal weight to the former two results.

On the other hand, a comparison with Ref.~\cite{xu2026effective} shows that the time parameterization of the dark matter equation of state may also have an important impact on the resulting constraints. A constant $w_{\rm dm}$ and a scale-factor-dependent parameterization, $w_{\rm dm}=w_2a^2$, describe different dark matter evolutionary histories and yield preferences for positive $w_{\rm dm}$ and negative $w_2$, respectively. This suggests that it will be necessary to test the current preference for a nonzero dark matter equation of state within more general frameworks allowing for time evolution.

Overall, although this work and several previous studies have found indications of a nonzero dark matter equation of state at the $2.6\sigma$--$3\sigma$ level in parameter posteriors, these results do not yet constitute a robust detection of non-cold dark matter. Future higher-precision observations of BAO, supernovae, weak lensing, and large-scale structure, together with joint analyses of the perturbative properties of dark energy and dark matter, will help to further test the current preference for a nonzero $w_{\rm dm}$ and clarify whether it represents new physics in the dark sector.

\section*{Acknowledgements}
This research was supported by the Research Startup funds of LongYan University (Grant No. LB2025037).



\bibliographystyle{spphys} 
\bibliography{references}






\end{document}